\documentclass[a4paper,fleqn,usenatbib]{mnras}

\usepackage{newtxtext,newtxmath}

\usepackage[T1]{fontenc}
\usepackage{ae,aecompl}

\usepackage{graphicx}	
\usepackage{amsmath}	
\usepackage{amssymb}	

\newcommand{\stwo}{S_{\rm 2}}
\newcommand{\sfour}{S_{\rm 4}}
\newcommand{\fourc}{S_{\rm 4C}}
\newcommand{\fourg}{S_{\rm 4G}}
\newcommand{\sixa}{S_{\rm 6A}}
\newcommand{\sixb}{S_{\rm 6B}}
\newcommand{\sixc}{S_{\rm 6C}}

\title[Symplectic Integrators: T + V Revisited]
{Symplectic Integrators: T + V Revisited and Round-Off Reduced}

\author[J. E. Chambers]{
John E. Chambers$^{1}$\thanks{E-mail: jchambers@carnegiescience.edu}
\\
$^{1}$Carnegie Institution for Science, 5241 Broad Branch Road NW, Washington DC 20015, USA\\
}

\date{Accepted XXX. Received YYY; in original form ZZZ}

\pubyear{2018}

\begin{document}
\label{firstpage}
\pagerange{\pageref{firstpage}--\pageref{lastpage}}
\maketitle

%
%
\begin{abstract}
Symplectic integrators separate a problem into parts that can be solved in isolation, alternately advancing these sub-problems to approximate the evolution of the complete system. Problems with a single, dominant mass can use mixed-variable symplectic (MVS) integrators that separate the problem into Keplerian motion of satellites about the primary, and satellite-satellite interactions. Here, we examine T+V algorithms where the problem is separated into kinetic $T$ and potential energy $V$ terms. T+V integrators are typically less efficient than MVS algorithms. This difference is reduced by using different step sizes for primary-satellite and satellite-satellite interactions. The T+V method is improved further using 4th and 6th-order algorithms that include force gradients and symplectic correctors. We describe three 6th-order algorithms, containing 2 or 3 force evaluations per step, that are competitive with MVS in some cases. Round-off errors for T+V integrators can be reduced by several orders of magnitude, at almost no computational cost, using a simple modification that keeps track of accumulated changes in the coordinates and momenta. This makes T+V algorithms desirable for long-term, high-accuracy calculations.
\end{abstract}


%
%
\section{Introduction} \label{sec:intro}
Symplectic integrators are often favored for integrating the orbits of planetary and satellite systems. These integrators have two important advantages over most other algorithms: (i) they show good long-term energy conservation properties, and (ii) they are computationally efficient for problems that involve a dominant central mass.

The usual strategy when devising a symplectic integrator is to divide the problem of interest into 2 or more parts that can each be solved easily in isolation. The algorithm then advances the parts alternately in a series of sub-steps that combine to approximate the evolution of the whole system. One possibility, that can be applied to any non-dissipative N-body problem, is to divide the problem into parts involving the kinetic energy $T$ and potential energy $V$ respectively \citep{gladman:1991}. This kind of algorithm is often referred to as a ``T+V'' integrator as a result. It is straightforward to advance each of these parts separately using Hamilton's equations. Either the momenta stay fixed while the coordinates change, or vice versa. 

T+V integrators are commonly used in some fields of physics, such as classical and quantum mechanical molecular dynamics problems \citep{bandrauk:1993, forbert:2001, omelyan:2002a}. However, their use for studying planetary and satellite systems in celestial mechanics has largely been superseded by ``mixed-variable'' symplectic (MVS) integrators \citep{wisdom:1991}. MVS algorithms take advantage of the presence of a dominant central body in the system, and split the problem into (i) Keplerian motion about the central body, and (ii) direct and indirect perturbations due to interactions between the less massive objects. This separation makes it possible to use substantially longer steps for the same level of accuracy as a T+V integrator.

One way to improve the efficiency of T+V algorithms is to use different step sizes for motion around the central body, and for other interactions. Since interactions between satellites are typically much weaker than the force from the central body, the satellite interactions can be advanced using a longer time step without compromising the accuracy of the algorithm. We will explore this possibility in this paper, and show that T+V integrators can be competitive with MVS algorithms in some circumstances, and that a simple modification makes them much more resistant to round-off errors.

The rest of this paper is organized as follows. Section 2 describes how  symplectic integrators can be constructed, and describes some examples that are accurate to second, fourth, and sixth-order in the step size. In Section~3, we apply these integrators to the 2-body Kepler problem, while Section~4 looks at systems containing more than two bodies. In Section~5, we show how T+V integrators can be modified to greatly reduce round-off errors. Section~6 contains a summary.

%
%
\section{Devising Symplectic Integrators}
Consider a system of $N$ bodies moving in 3 spatial dimensions described by a Hamiltonian $H_F$. Using Hamilton's equations, the evolution of any quantity $q$ can be expressed as
\begin{eqnarray}
\frac{dq}{dt}&=&
\sum_{i=1}^{3N}
\left(\frac{dx_i}{dt}\frac{\partial q}{\partial x_i}
+\frac{dp_i}{dt}\frac{\partial q}{\partial p_i}\right)
\nonumber \\
&=&\sum_{i=1}^{3N}
\left(\frac{\partial H_F}{\partial p_i}\frac{\partial }{\partial x_i}
-\frac{\partial H_F}{\partial x_i}\frac{\partial }{\partial p_i}\right)q
\nonumber \\
&=&Fq
\end{eqnarray}
where $t$ is the time, and ${\bf x}_i$ and ${\bf p}_i$ are coordinates and momenta of body $i$. Here $F$ is an operator that encapsulates the evolution of the system and depends on $H_F$.

The value of $q$ after one time step $\tau$ can be expressed as
\begin{equation}
q(\tau)=e^{\tau F}q(0)=\left(1+\tau F+\frac{\tau^2}{2}F^2+\cdots
\right)q(0)
\end{equation}

For most $N$-body systems, $F$ is too complicated to allow an exact solution. However, as we noted in the introduction, we can usually separate the problem into 2 or more parts that {\em are\/} easy to solve. For example if we separate the Hamiltonian so that $H_F=H_A+H_B$, with corresponding operators $A$ and $B$, then $q(\tau)$ is given by
\begin{equation}
q(\tau)=e^{\tau(A+B)}q(0)
\end{equation}
where
\begin{eqnarray}
e^{\tau(A+B)}&=&
1+\tau(A+B)+\frac{\tau^2}{2}(A+B)^2+\cdots
\nonumber \\
&=&1+\tau(A+B)+\frac{\tau^2}{2}(A^2+AB+BA+B^2)+\cdots
\nonumber \\
\label{eq_full_system}
\end{eqnarray}
We note that the operators $A$ and $B$ do not commute in general, so that $AB\neq BA$.

A simple integrator consists of just two steps in which the system is advanced for one time step under each of the two sub-problems separately. Using the operators $A$ and $B$, we can describe this integrator as
\begin{equation}
e^{\tau A}e^{\tau B}=
\exp\left\{
\tau(A+B)+\frac{\tau^2}{2}[A,B]+\cdots
\right\}
\label{eq_first_order}
\end{equation}
where the square brackets indicate a commutator defined by $[A,B]=AB-BA$. 

The algorithm described by Eqn.~\ref{eq_first_order} differs from the true system, described by Eqn.~\ref{eq_full_system} by a factor $O(\tau^2)$, so the integrator is accurate to first order in the step size.

%
%
\subsection{Second and 4th-Order Integrators}
We can devise higher-order integrators using the Baker-Campbell-Hausdorff formula \citep{yoshida:1990}, which gives the product of exponential operators in terms of a series of commutators:
\begin{eqnarray}
e^Ae^B&=&\exp\left(A+B+\frac{1}{2}[A,B]+\frac{1}{12}[A,A,B]+\frac{1}{12}[B,B,A]
\right.
\nonumber \\
&+&\left.\frac{1}{24}[A,B,B,A]+\cdots\right)
\end{eqnarray}
where we use Yoshida's compact commutator notation $[A,B,C]\equiv[A,[B,C]]$ and $[A,B,C,D]\equiv[A,[B,[C,D]]]$ etc.

The well-known second-order leapfrog integrator consists of 3 sub-steps:
\begin{equation}
S_2=e^{\tau B/2}e^{\tau A}e^{\tau B/2}
=\exp\left\{\tau(A+B)+O(\tau^3)+\cdots
\right\}
\label{eq_s2}
\end{equation}
Using this algorithm, the sub-system $H_B$ is advanced for half a time step, then the sub-system $H_A$ is advanced for a full step, followed by another half step for sub-system $H_B$.

Leapfrog is an example of a time-symmetric integrator where the algorithm is unchanged if the sequence of sub-steps is reversed. Symmetric algorithms have the advantage that they contain no error terms with even powers of the timestep $\tau$ \citep{yoshida:1990}, and we will only consider symmetric integrators from now on.

\cite{forest:1990} described a 4th-order algorithm that contains 7 sub-steps:
\begin{eqnarray}
S_4&=&\exp(a\tau B)
\exp(2a\tau A)
\exp\left\{\left(\frac{1}{2}-a\right)\tau B\right\}
\exp\left\{(1-4a)\tau A\right\}
\nonumber \\
&&\exp\left\{\left(\frac{1}{2}-a\right)\tau B\right\}
\exp(2a\tau A)
\exp(a\tau B)
\nonumber \\
&=&\exp\left\{\tau(A+B)+O(\tau^5)+\cdots\right\}
\label{eq_s4}
\end{eqnarray}
where $a=1/(4-2^{4/3})=0.6756\ldots$. Although this integrator is accurate to 4th-order in the step size, it does not perform as well as one might expect since the sub-steps are large (and some travel backwards in time), so the coefficients of the leading error terms are large \citep{chambers:2003}.

%
%
\subsection{Force Gradients}
One way to reduce the size of the sub-steps is to use terms with ``force gradients'' \citep{omelyan:2002b}. Many $N$-body problems contain only quadratic momentum terms.
If we split the Hamiltonian into two parts such that one of them $H_A$ contains all of the momentum terms, and the other $H_B$ depends only on coordinates, then the operator $[B,B,A]$ will depend on coordinates only. A sub-step consisting of the $[B,B,A]$ operator can be advanced easily according to Hamilton's equations since the coordinates remain fixed for the duration of that sub-step.

A simple 4th-order integrator that includes a force-gradient term is
\begin{eqnarray}
S_{4G}&=&\exp\left(\frac{\tau B}{6}\right)
\exp\left(\frac{\tau A}{2}\right)
\exp\left(\frac{2\tau B}{3}-[B,B,A]\frac{\tau^3}{72}\right)
\nonumber \\
&&\exp\left(\frac{\tau A}{2}\right)
\exp\left(\frac{\tau B}{6}\right)
\nonumber \\
&=&\exp\left\{\tau(A+B)+O(\tau^5)+\cdots\right\}
\label{eq_s4g}
\end{eqnarray}
Note that the $[B,B,A]$ force gradient term can be advanced at the same time as the $2\tau B/3$ term since both these sub-steps only alter the momenta. We will examine the form of $[B,B,A]$ in more detail in Sections~3 and 4.

%
%
\subsection{Symplectic Correctors}
Another strategy for developing integrators is to use a ``symplectic corrector'' \citep{wisdom:1996}. Correctors are a sequence of sub-steps applied before and after an integration step such that the terms at the end of one step exactly cancel those applied at the start of the next step. In practice, this means the corrector or its inverse needs only be applied at the start of an integration and immediately before output is required. Correctors typically consist of a complicated series of sub-steps, and can be computationally expensive. However, they are worth the cost if output is not needed very often.

\citet{wisdom:1996} provide a useful formula for the effect of a corrector $C$ on an unmodified integrator or ``kernel'' $K$:
\begin{equation}
e^Ce^Ke^{-C}=
\exp\left(K+[C,K]+\frac{1}{2}[C,C,K]+\frac{1}{6}[C,C,C,K]+\cdots
\right)
\end{equation}

Using this formula, we can obtain the following 4th-order integrator that includes a force gradient and a corrector:
\begin{eqnarray}
S_{4C}&=&
\exp\left(\frac{\tau^2}{12}[A,B]\right)
\exp\left(\frac{\tau B}{2}-\frac{\tau^3}{48}[B,B,A]\right)
\exp(\tau A)
\nonumber \\
&&\exp\left(\frac{\tau B}{2}-\frac{\tau^3}{48}[B,B,A]\right)
\exp\left(-\frac{\tau^2}{12}[A,B]\right)
\nonumber \\
&=&\exp\left\{\tau(A+B)+O(\tau^5)+\cdots\right\}
\label{eq_s4c}
\end{eqnarray}

In general, we can only get an approximate expression for $[A,B]$ accurate to some order in the step size. The following corrector is accurate to $O(\tau^3)$, which is sufficient to make the whole integrator 4th-order accurate:
\begin{eqnarray}
&&\exp\left(\frac{\tau^2}{12}[A,B]\right)\simeq
\nonumber \\
&&\exp\left(\frac{\tau A}{4}\right)
\exp\left(\frac{\tau B}{6}\right)
\exp\left(-\frac{\tau A}{4}\right)
\exp\left(-\frac{\tau B}{6}\right)
\exp\left(-\frac{\tau A}{4}\right)
\nonumber \\
&&\exp\left(-\frac{\tau B}{6}\right)
\exp\left(\frac{\tau A}{4}\right)
\exp\left(\frac{\tau B}{6}\right)
\end{eqnarray}
The inverse corrector reverses the order of these sub-steps as well as their signs. 

%
%
\subsection{Sixth-Order Algorithms}
Sixth order symplectic integrators typically require many substeps. For example, \citet{yoshida:1990} gives three examples that each contain 15 substeps. The number of sub-steps can be reduced substantially when force gradients and symplectic correctors are included. In this subsection, we will examine three sixth-order algorithms, one that consists of 5 sub-steps, and two that contain 7 sub-steps. All of these use ordinary force-gradient terms as well as a higher-order derivative of the force gradient that is proportional to the operator $[B,B,A,A,B]$. For systems with quadratic momenta, this operator consists only of coordinates, and can be advanced in the same way as $[B,B,A]$.

The following sixth-order algorithm contains 5 sub-steps plus a corrector:
\begin{eqnarray}
S_{6A}&=&\exp(j\tau^2[A,B]+k\tau^4[A,A,A,B]+l\tau^4[A,B,B,A])
\nonumber \\
&&\exp(a\tau A)
\exp\left(\frac{\tau B}{2}+g\tau^3[B,B,A]+h\tau^5[B,B,A,A,B]\right)
\nonumber \\
&&\exp\left\{(1-2a)\tau A\right\}
\nonumber \\
&&\exp\left(\frac{\tau B}{2}+g\tau^3[B,B,A]+h\tau^3[B,B,A,A,B]\right)
\exp(a\tau A)\
\nonumber \\
&&\exp(-j\tau^2[A,B]-k\tau^4[A,A,A,B]-l\tau^4[A,B,B,A])
\nonumber \\
&=&\exp\left(\tau(A+B)+O(\tau^7)+\cdots\right)
\label{eq_s6a}
\end{eqnarray}
where
\begin{eqnarray}
a&=&\frac{1}{4}+\frac{1}{4}\left(1+\frac{4}{\surd 15}\right)^{1/2}
\simeq0.606,440,349,058,282\ldots
\nonumber \\
g&=&-\frac{1}{48}+\frac{a}{8}-\frac{a^2}{4}
\simeq-0.036,970,763,942,530\ldots
\nonumber \\
h&=&\frac{1}{2880}-\frac{a}{96}+\frac{a^2}{12}-\frac{a^3}{4}+\frac{a^4}{4}
\simeq0.002,733,674,772,988\ldots
\nonumber \\
j&=&\frac{1}{12}-\frac{a}{2}+\frac{a^2}{2}
\simeq-0.036,001,892,712,842\ldots
\nonumber \\
k&=&-\frac{1}{720}+\frac{a^2}{24}-\frac{a^3}{12}+\frac{a^4}{24}
\simeq0.000,984,593,807,026\ldots
\nonumber \\
l&=&\frac{1}{720}+\frac{a}{48}-\frac{5a^2}{24}+\frac{a^3}{2}-\frac{3a^4}{8}
\nonumber \\
&\simeq&-0.001,800,924,780,266\ldots
\end{eqnarray}

This integrator is accurate to sixth order in the step size provided that the Hamiltonian for the system contains only quadratic momenta, in which case $[B,B,B,A]=[A,B,B,B,A]=[B,B,B,B,A]=0$. Note that unlike the previous algorithms, the kernel of this integrator begins with a sub-step involving operator $A$ rather than $B$. There is no equivalent 6th-order algorithm beginning with $B$ that has sub-steps with real coefficients.

Two other 6th-order algorithms will be useful later due to the special form of the corrector in each case. These additional constraints on the correctors mean that each integrator kernel requires 7 sub-steps instead of 5.

The first 7-step integrator has the following form
\begin{eqnarray}
S_{6B}&=&\exp(k\tau^4[A,A,A,B]+l\tau^4[A,B,B,A])
\nonumber \\
&&\exp\left\{b\tau B+g\tau^3[B,B,A]+h\tau^5[B,B,A,A,B]\right\}
\nonumber \\
&&\exp(a\tau A)\exp\left\{(\frac{1}{2}-b)\tau B\right\}
\exp\left\{(1-2a)\tau A\right\}
\nonumber \\
&&\exp\left\{(\frac{1}{2}-b)\tau B\right\}\exp(a\tau A)
\nonumber \\
&&\exp\left\{b\tau B+g\tau^3[B,B,A]+h\tau^3[B,B,A,A,B]\right\}
\nonumber \\
&&\exp(-k\tau^4[A,A,A,B]-l\tau^4[A,B,B,A])
\label{eq_s6b}
\end{eqnarray}
where $a$ is the smaller real root of 
\begin{equation}
30a^4-90a^3+78a^2-26a+3=0
\end{equation}
which gives $a\simeq0.577,953,138,043,435\cdots$, and
\begin{eqnarray}
b&=&\frac{(6a^2-6a+1)}{12a(a-1)}\simeq0.158,362,565,165,888\cdots
\nonumber \\
g&=&\frac{(6a^3-12a^2+6a-1)}{288a(a-1)^2}\simeq-0.012,894,895,451,727\cdots
\nonumber \\
k&=&-\frac{(5a^2-5a+1)}{720}\simeq0.000,305,022,974,091\cdots
\nonumber \\
l&=&\frac{(6a^2-2a+1)}{2880(a-1)^2}\simeq-0.003,602,900,019,507\cdots
\nonumber \\
h&\simeq&-0.000,486,709,920,391\cdots
\end{eqnarray}

Another 7-step integrator has the following form (note the different location of the $g$ and $h$ terms):
\begin{eqnarray}
S_{6C}&=&\exp(j\tau^2[A,B]+l\tau^4[A,B,B,A])
\nonumber \\
&&\exp(b\tau B)\exp(a\tau A)
\nonumber \\
&&\exp\left\{(\frac{1}{2}-b)\tau B+g\tau^3[B,B,A]+h\tau^5[B,B,A,A,B]\right\}
\nonumber \\
&&\exp\left\{(1-2a)\tau A\right\}
\nonumber \\
&&\exp\left\{(\frac{1}{2}-b)\tau B+g\tau^3[B,B,A]+h\tau^3[B,B,A,A,B]\right\}
\nonumber \\
&&\exp(a\tau A)\exp(b\tau B)
\nonumber \\
&&\exp(-j\tau^2[A,B]-l\tau^4[A,B,B,A])
\label{eq_s6c}
\end{eqnarray}
where $a$ is the real root of 
\begin{equation}
15a^5-60a^4+90a^3-60a^2+18a-2=0
\end{equation}
which gives $a\simeq0.567,040,718,865,478\cdots$, and
\begin{eqnarray}
b&=&\frac{(30a^4-60a^3+30a^2-1)}{60a^2(a-1)^2}
\simeq0.223,480,254,150,115\cdots
\nonumber \\
g&=&-\frac{1}{48}-\frac{1}{120a(a-1)}-\frac{1}{7200a^3(a-1)^4}
\nonumber \\
&\simeq&-0.008,568,633,689,896\cdots
\nonumber \\
j&=&\frac{(5a^2-5a+1)}{60a(a-1)}
\simeq0.015,446,203,250,883\cdots
\nonumber \\
l&=&-\frac{1}{144}-\frac{a(a-1)}{48}
+\frac{(1-2a^2)}{14400a^2(a-1)^4}
\nonumber \\
&\simeq&0.000,364,086,621,888\cdots
\nonumber \\
h&\simeq&0.000,241,417,111,491\cdots
\end{eqnarray}

For each of the 6th-order algorithms, we need a corrector composed of alternating $A$ and $B$ operators that is accurate to $O(\tau^5)$. These correctors contain only even powers of $\tau$. We note that a corrector of the following form can be adapted to give any corrector with even powers of $\tau$ up to 4th-order by choosing appropriate values of the free parameters $\alpha_{1,2}$ and $\beta_{1,2}$:
\begin{eqnarray}
e^C&=&\prod_{i=1}^{2}
\exp(X_i)
\exp-(X_i)
\exp(-X_i)
\exp(X_i)
\exp(-X_i)
\nonumber \\
&&\exp(-X_i)
\exp(X_i)
\exp(-X_i)
\end{eqnarray}
where
\begin{equation}
\exp(X_i)=\exp(\alpha_i\tau A)\exp(\beta_i\tau B)
\label{eq_x}
\end{equation}

These correctors contain 32 terms and are expensive to calculate, but the resulting integrators remain efficient as long as output is not required too often. (It is possible that the desired correctors could be constructed with fewer terms, but we do not explore this here.)

The values of the $\alpha$ and $\beta$ are related to the coefficients $j$, $k$ and $l$ of the $[A,B]$, $[A,A,A,B]$ and $[A,B,B,A]$ terms in the corrector by the following equations:
\begin{eqnarray}
j&=&4(\alpha_1\beta_1+\alpha_2\beta_2) \nonumber \\
k&=&\frac{2}{3}\left(\alpha_1^3\beta_1+\alpha_2^3\beta_2\right) \nonumber \\
l&=&-(\alpha_1^2\beta_1^2+\alpha_2^2\beta_2^2)
\end{eqnarray}
Note that the sign of the $l$ term can be changed by reversing the order of $A$ and $B$ in Eqn.~\ref{eq_x}

%
%
\section{The Kepler Problem}
In this section, we apply the integration algorithms described above to the Kepler problem in which a test particle orbits a point mass $M$. The Hamiltonian in this case is
\begin{equation}
H_F=\frac{p^2}{2}-\frac{GM}{r}
\end{equation}
where $r$ is the distance between the objects.

We separate the Hamiltonian into kinetic and potential energy terms:
\begin{eqnarray}
H_A&=&\frac{p^2}{2}
\nonumber \\
H_B&=&-\frac{GM}{r}
\end{eqnarray}

With this separation, $H_A$ can be advanced by keeping the momenta fixed and linearly increasing the positions at constant velocity. Similarly, $H_B$ can be advanced by keeping the coordinates fixed and applying a constant acceleration to the velocities. The force gradient operators used by some of the higher-order integrators are particularly simple in this case:
\begin{eqnarray}
\tau^3[B,B,A]&=&\frac{G^2M^2\tau^3}{r^4}
\nonumber \\
\tau^5[B,B,A,A,B]&=&-\frac{4G^3M^3\tau^5}{r^7}
\end{eqnarray}
and it is straightforward to advance the system under either of these operators.

\begin{figure}
\includegraphics[width=\columnwidth]{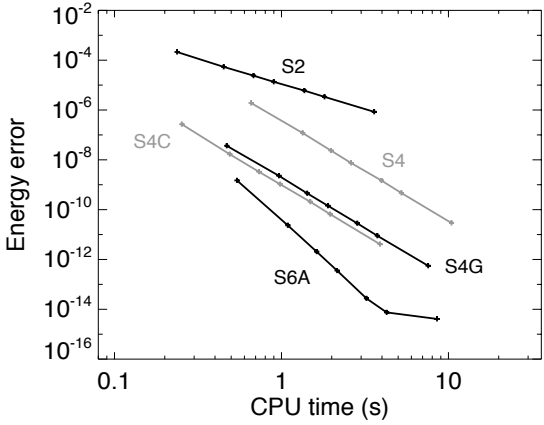}
\caption{Root-mean-squared energy error versus computation time for integrations of the two-body Kepler problem. Each curve shows calculations using a particular integrator for various step sizes. $\stwo$ and $\sfour$ denote classical 2nd and 4th-order T+V algorithms,  given by Eqns.~\ref{eq_s2} and \ref{eq_s4}. $\fourg$ is a 4th-order algorithm using a force gradient, given by Eqn.~\ref{eq_s4g}. $\fourc$ and $\sixa$ are 4th and 6th-order algorithms that use force gradients and symplectic correctors, given by Eqns.~\ref{eq_s4c} and \ref{eq_s6a}.}
\end{figure}

Figure 1 shows the performance of some of the T+V integrators described above when applied to the Kepler problem. The figure shows the rms energy error as a function of computer time for integrations lasting 100,000 orbital periods for an orbit with an eccentricity of 0.1. From the slope of the curves, we see that the 2nd, 4th and 6th-order algorithms are behaving as expected. For example, decreasing the step size by a factor of 2 doubles the computation time while roughly reducing the error by factors of 4, 16 and 64 respectively. The error for the 6th-order integrator levels off at very small step sizes due to round-off error.

%
%
\section{The $N$-Body Problem}
In this section, we consider the evolution of a system containing several bodies orbiting a dominant central mass. We will work with democratic heliocentric coordinates \citep{duncan:1998}, which consist of coordinates ${\bf X}$ with respect to the central body, and momenta ${\bf P}$ with respect to the center of mass. Using these coordinates, the Hamiltonian can be split into the following parts:
\begin{eqnarray}
H_A&=&\sum_{i=1}^N\frac{P_i^2}{2m_i}+\left(\sum\frac{{\bf P}_i}{2m_0}\right)^2
\nonumber \\
H_B&=&-\sum_{i=1}^N\frac{Gm_0m_i}{R_{i0}}
\nonumber \\
H_I&=&-\sum_{i=1}^N\sum_{j>i}\frac{Gm_im_j}{R_{ij}}
\end{eqnarray}
where $N$ is the number of satellites orbiting the central body which has index 0. 

Note that $H_I$ is typically much smaller than $H_A$ and $H_B$ for small satellite-to-primary mass ratios. We can make use of this difference to produce a modified leapfrog algorithm with the following steps
\begin{itemize}
\item Apply a corrector $C_I$.
\item Advance $H_I$ for $\tau/2$
\item Do the following $M$ times:
\begin{itemize}
\item Advance $H_B$ for $\tau/(2M)$
\item Advance $H_A$ for $\tau/M$
\item Advance $H_B$ for $\tau/(2M)$
\end{itemize}
\item Advance $H_I$ for $\tau/2$
\item Apply an inverse corrector $-C_I$.
\end{itemize}
where $M$ is an integer and $\tau$ is the step size. When $M=1$, we have the usual leapfrog algorithm. More efficient algorithms will use $M>1$. The speed-up can be significant when $N$ is larger than a few, since advancing $H_I$ requires $O(N^2)$ operations compared with $O(N)$ for $H_A$ and $H_B$.

Note that we have included a corrector $C_I$ to eliminate the leading error term that contains a single factor of $I$, where
\begin{equation}
C_I=\frac{\tau^2}{12}[A,I]
\end{equation}
We do not need to include $B$ here since $[B,I]=0$. This corrector doesn't do much to improve the performance of leapfrog, but it will improve some of the higher-order algorithms discussed below, so we include it here for consistency.

We can use a similar procedure for the conventional 4th-order integrator $S_4$. The 4th-order gradient integrator $S_{4G}$ requires minor modification since the gradient is more complicated than for the Kepler problem of the previous section. The 4th-order gradient algorithm has the following steps:
\begin{itemize}
\item Apply a corrector $C_I$.
\item Advance $H_I$ for $\tau/2$
\item Do the following $M$ times:
\begin{itemize}
\item Advance $H_B$ for $\tau/(6M)$
\item Advance $H_A$ for $\tau/(2M)$
\item Advance $H_B$ for $2\tau/(3M)$ and $[B,B,A]$ for $-\tau^3/(72M)$
\item Advance $H_A$ for $\tau/(2M)$
\item Advance $H_B$ for $\tau/(6M)$
\end{itemize}
\item Advance $H_I$ for $\tau/2$
\item Apply an inverse corrector $-C_I$.
\end{itemize}
where the gradient operator is now given by
\begin{equation}
\tau^3[B,B,A]=G^2m_0\tau^3\left(
\sum_{i=1}^N\frac{m_0m_i}{R_i^4}+Q^2\right)
\end{equation}
where
\begin{equation}
{\bf Q}=\sum_{i=1}^N\frac{m_i{\bf R}_i}{R_i^3}
\end{equation}

\begin{figure}
\includegraphics[width=\columnwidth]{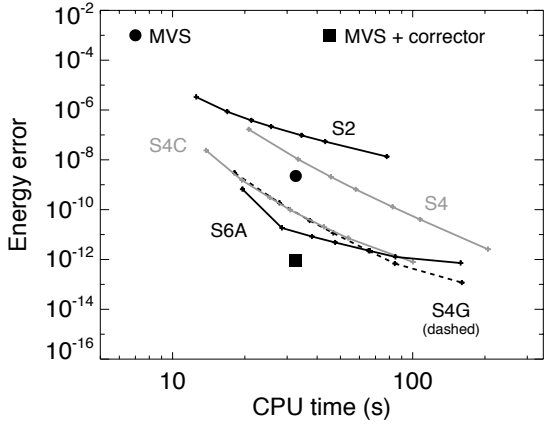}
\caption{Root-mean-squared energy error versus computation time for integrations of the Sun and 8 planets of the Solar System. Each curve shows calculations using a particular integrator for various step sizes. The labels $\stwo$ etc have the same meaning as in Figure~1. The circle and square symbols show results for a second-order MVS integrator, and a second-order MVS with a symplectic corrector, respectively.}
\end{figure}

Figure 2 shows the performance of the T+V integrators when integrating the Sun and the 8 planets of the Solar System. The integrations last for 100,000 years. The step size for the direct terms $H_I$ is 1.8 days, while cases for multiple values of $M$ are shown. The figure also shows integrations using a 2nd-order leapfrog MVS integrator, and a 2nd-order MVS leapfrog with a corrector that is included with the {\em Mercury\/} $N$-body integration package \citep{chambers:1999, chambers:2010}. For the MVS integrators, we advance the Keplerian orbits using the accurate and efficient routine described by \citet{rein:2015}.

The classical second and fourth order integrators $S_2$ and $S_4$ perform poorly compared to the MVS integrator. However, the other fourth order algorithms and the sixth order integrator $\sixa$ are more efficient (requiring less CPU time for a given accuracy) than the standard MVS algorithm. MVS with a corrector is more efficient than all the T+V integrators, although it is only slightly better than $\sixa$. Thus, the usual great speed disadvantage of T+V  compared to MVS integrators can be substantially reduced by using these 4th and 6th order algorithms.

The accuracy of the second and fourth order integrators varies with the step size roughly as expected. Doubling the step size roughly doubles the integration cost, while reducing the error by factors of about 4 and 16 for the second and fourth order cases respectively. However, $\sixa$ doesn't perform as well as expected. In fact, the slope of the curve in Figure~2 shows that $\sixa$ mostly behaves as a 2nd order integrator rather than 6th order. In the following subsections we explore the reason for this behavior and show how it can be fixed.

%
%
\subsection{Improved Sixth-Order Integrators}
We can get a sense of the problem with the 6th-order algorithm $\sixa$ by examining the simpler 4th-order algorithm, $\fourc$, described by Eqn.~\ref{eq_s4c}, with the inclusion of direct terms that are advanced by the operator $I$. This algorithm also uses a corrector. Consider the simplest case in which the step size for $I$ is the same as for $A$ and $B$. The integrator kernel $K$ is then
\begin{eqnarray}
\exp(K)&=&
\exp\left(\frac{\tau I}{2}\right)
\exp\left(\frac{\tau B}{2}-\frac{\tau^3}{48}[B,B,A]\right)
\exp(\tau A)
\nonumber \\
&&\exp\left(\frac{\tau B}{2}-\frac{\tau^3}{48}[B,B,A]\right)
\exp\left(\frac{\tau I}{2}\right)
\label{eq_kernel_s4c}
\end{eqnarray}

We then want to apply a corrector $C$ outside the kernel. The most general corrector available at second order in the step size has the form
\begin{equation}
C=j\tau^2[A,B]+k\tau^2[A,I]+O(\tau^4)
\end{equation}
where $j$ and $k$ are constants. Note that $[B,I]=0$ since both the corresponding pieces of the Hamiltonian depend only on the coordinates. Using this general corrector, a complete step of the integrator is
\begin{eqnarray}
&&\exp(C)\exp(K)\exp(-C)=
\exp\left\{\tau(A+B+I)\right.
\nonumber \\
&&\left.+\left(\frac{1}{12}-j\right)\tau^3[A,A,B]
+\left(j-\frac{1}{12}\right)\tau^3[B,B,A]
\right.
\nonumber \\
&&\left.+\left(\frac{1}{12}-k\right)\tau^3[A,A,I]
+\left(\frac{1}{12}-j-k\right)\tau^3[B,A,I]\right.
\nonumber \\
&&\left.+\left(k-\frac{1}{24}\right)\tau^3[I,I,A]+O(\tau^5)+\cdots\right\}
\end{eqnarray}

The $[A,A,B]$ and $[B,B,A]$ terms are both eliminated by choosing $j=1/12$ as in the original S4C integrator given in Eqn.~\ref{eq_s4c}. The $[I,I,A]$ term can be neglected since it contains 2 factors of the small quantity $I$. However, it is impossible to eliminate both the remaining $O(\tau^3)$ terms with a single value of $k$. This means that, in principle, the integrator will behave as second order rather than 4th-order.

\begin{figure}
\includegraphics[width=\columnwidth]{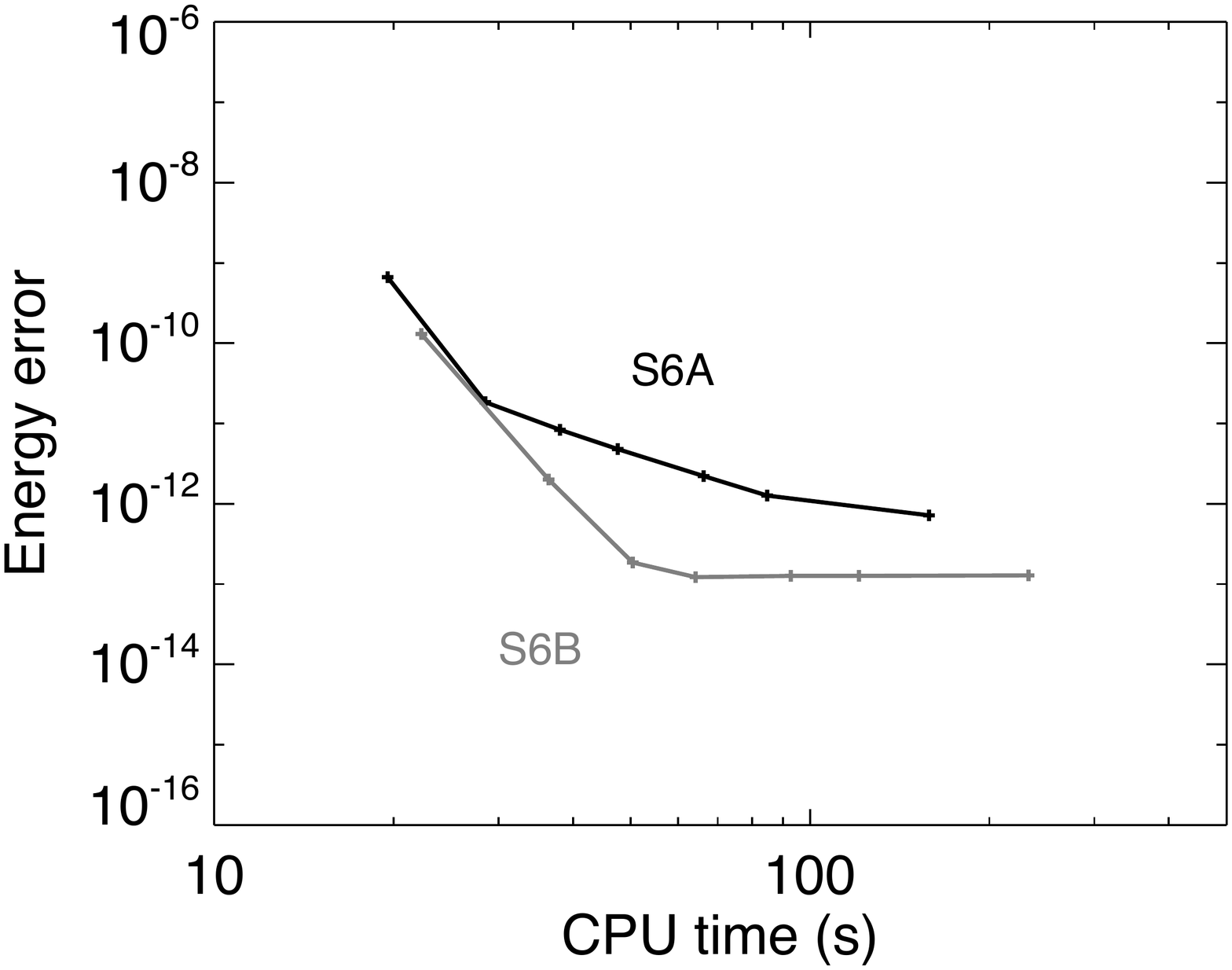}
\caption{Root-mean-squared energy error versus computation time for integrations of the Sun and 8 planets of the Solar System. Each curve shows calculations using a particular integrator for various step sizes. $\sixa$ and $\sixb$ are different 6th-order algorithms given by Eqns.~\ref{eq_s6a} and \ref{eq_s6b} respectively.}
\end{figure}

The underlying cause of the problem is that the $[A,B]$ term in the corrector should really be applied in between the sub-steps involving $I$ in Eqn.~\ref{eq_kernel_s4c} rather than outside them. However, this would entail applying the corrector and its inverse at the same frequency as the $I$ sub-steps, which would be inefficient since the corrector is typically expensive. The results shown in Figure 2 suggest that the error incurred by moving the corrector outside the $I$ steps is not a problem for the 4th-order integrator $S_{4C}$, presumably because the uncorrected terms contain a factor of $I$ which is small. However, the problem becomes obvious for the 6th-order algorithm, since multiple terms at $O(\tau^3)$ and $O(\tau^5)$ are not properly eliminated when the corrector is moved.

One way to overcome this problem is to use an integrator kernel that doesn't require an $[A,B]$ term in its corrector. For example, if the integrator kernel without the $I$ steps has the following form
\begin{equation}
\exp(K)=\exp\{\tau(A+B)+O(\tau^5)+\cdots\}
\end{equation}
then it will have the following form when the $I$ steps are added
\begin{eqnarray}
&&\exp\left(\frac{I}{2}\right)\exp(K)\exp\left(\frac{I}{2}\right)
=\exp\left\{\tau(A+B+I)
+\frac{\tau^3}{12}[A,A,I]\right.
\nonumber \\
&&\left.+\frac{\tau^3}{12}[B,A,I]
-\frac{\tau^3}{24}[I,I,A]+\cdots
\right\}
\end{eqnarray}
so that both the $[A,A,I]$ and $[B,A,I]$ terms can be eliminated by a corrector of the form $(\tau^2/12)[A,I]$. Some terms at $O(\tau^5)$ will remain uncorrected regardless of what other terms are in the corrector, but these uncorrected terms will contain at least one factor of $I$, and should be small enough not to degrade the integrator's performance. 

The integrator $\sixb$, described by Eqn.~\ref{eq_s6b} in Section 2.4 has a corrector with the necessary properties, and we now examine the performance of this integrator. Figure~3 shows the performance of $\sixb$ compared to $\sixa$ for an integration of the Sun and 8 planets of the Solar System. The integrations last for 100,000 years. The step size for the direct planet-planet terms, represented by $H_I$, is held constant at 1.8 days, while several values of the step size are considered for the other parts of the Hamiltonian.

For large step sizes, the energy error of $\sixb$ varies as roughly as the sixth power of the step size, so the algorithm is indeed behaving as a sixth-order integrator, as desired. The computational cost is generally smaller than for $\sixa$ for a given energy error. As we saw earlier, $\sixa$ mostly behaves as a second-order algorithm. For small step sizes, the accuracy of $\sixb$ saturates at about 1 part in $10^{-13}$. Further reductions in the step size do not improve the accuracy. This implies that the dominant source of error at this point is the direct planet-planet terms of $H_I$. Reducing the error further would require reducing the step size for these terms as well as the other parts of the Hamiltonian.

%
%
\subsection{Simple, Exact Correctors}
Another strategy for improving the 6th-order integrator $\sixa$ is to return the corrector to its correct position and find a way to implement the corrector more efficiently. This means that the corrector can be applied at every integration step as it should be. It turns out that this can be done, at least for the Kepler problem where the Hamiltonian equivalent to the operator $\tau^2[A,B]$ is integrable and can be solved efficiently.

For the Kepler problem, we have
\begin{eqnarray}
H_A&=&\frac{p^2}{2}
\nonumber \\
H_B&=&-\frac{GM}{r}
\end{eqnarray}
and the Hamiltonian equivalent to $\tau^2[A,B]$ is
\begin{eqnarray}
H_{AB}&=&\tau\left(\frac{\partial H_A}{\partial x}\frac{\partial H_B}{\partial p_x}
-\frac{\partial H_A}{\partial p_x}\frac{\partial H_B}{\partial x}\right)
+{\rm (y, z\ terms)}
\nonumber \\
&=&
-\frac{GM\tau}{r^3}({\bf x}\cdot{\bf p})
\end{eqnarray}

Advancing the system under $H_{AB}$ using Hamilton's equation, we get
\begin{eqnarray}
\frac{dx}{dt}&=&\frac{\partial H_{AB}}{\partial p_x}
=-\frac{GM\tau x}{r^3}
\nonumber \\
\frac{dp_x}{dt}&=&-\frac{\partial H_{AB}}{\partial x}
=\frac{GM\tau p_x}{r^3}
-\frac{3GM\tau x}{r^5}({\bf x}\cdot{\bf p})
\end{eqnarray}

Noting that the evolution of the coordinates does not depend on the momenta, we can solve these equations analytically to get
\begin{eqnarray}
r^3(t)&=&r_0^3-3GM\tau t \nonumber \\
{\bf x}(t)&=&{\bf x}_0\frac{r(t)}{r_0} \nonumber \\
{\bf p}(t)&=&\frac{r_0}{r(t)}\left\{{\bf p}_0
-\frac{3GM\tau t}{r_0^5}({\bf x}_0\cdot {\bf p}_0){\bf x}_0
\right\}
\end{eqnarray}
where the subscript 0 indicates the initial values, and we have used the fact that $H_{AB}$ is a constant to solve the momentum equations. 

A corrector with the form $\tau^4[A,B,B,A]$ can be advanced analytically in a similar way, but it is not obvious how to do the same for $\tau^4[A,A,A,B]$. Therefore, we will use integrator $\sixc$, described by Eqn.~\ref{eq_s6c} in Section 2.4, which has a corrector that only contains terms that we can advance analytically.

To use this integrator, we once again adopt democratic heliocentric coordinates, but separate out the indirect momentum terms, so that the Hamiltonian has the following parts:
\begin{eqnarray}
H_A&=&\sum_{i=1}^N\frac{P_i^2}{2m_i}
\nonumber \\
H_B&=&-\sum_{i=1}^N\frac{Gm_0m_i}{R_{i0}}
\nonumber \\
H_S&=&\left(\sum\frac{{\bf P}_i}{2m_0}\right)^2
\nonumber \\
H_I&=&-\sum_{i=1}^N\sum_{j>i}\frac{Gm_im_j}{R_{ij}}
\end{eqnarray}
where $N$ is the number of satellites orbiting the central body which has index 0. Note that $H_A$ and $H_B$ correspond to $N$ separate Kepler problems, so we can use the analytic corrector derived above. 

One step of the integrator looks like this:
\begin{itemize}
\item Apply a corrector $C_I$
\item Advance $H_I$ for $\tau/2$
\item Advance $H_S$ for $\tau/2$
\item Apply an analytic corrector proportional to $\tau^2[A,B]$
\item Apply an analytic corrector proportional to $\tau^4[A,B,B,A]$
\item Do the following $M$ times
\begin{itemize}
\item Apply the 7-step kernel of integrator $S_{\rm 6C}$ applied to $H_A$ and $H_B$ only.
\end{itemize} 
\item Apply an analytic inverse corrector proportional to $\tau^4[A,B,B,A]$
\item Apply an analytic inverse corrector proportional to $\tau^2[A,B]$
\item Advance $H_S$ for $\tau/2$
\item Advance $H_I$ for $\tau/2$
\item Apply an inverse corrector $-C_I$
\end{itemize}
where we have included the corrector $C_I$ to eliminate the leading error term involving $I$. This needs only be applied at the start of the integration and when output is required. We note that the $[A,B]$ and $[A,B,B,A]$ correctors can be applied one after the other without appreciably affecting the accuracy. 

\begin{figure}
\includegraphics[width=\columnwidth]{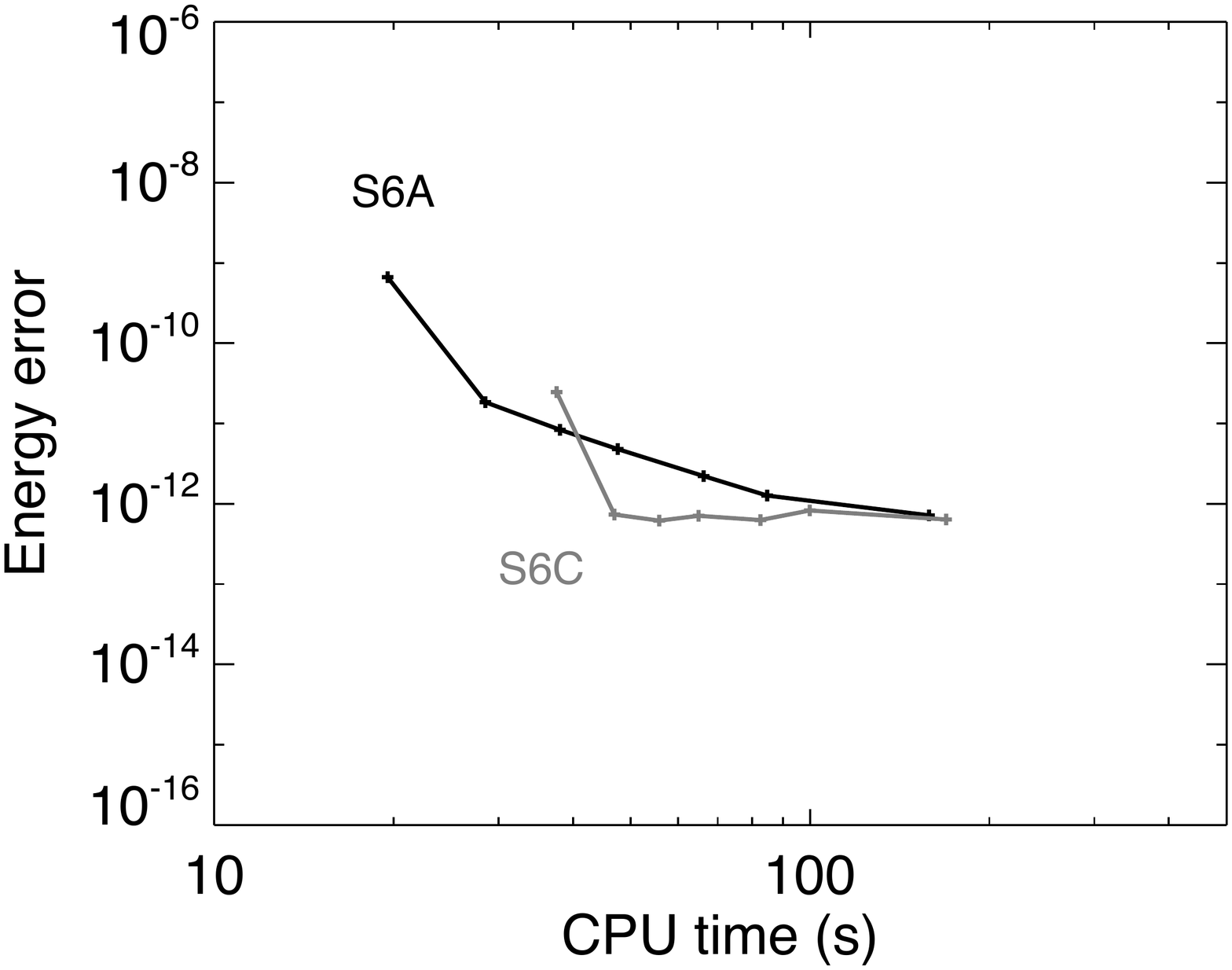}
\caption{Root-mean-squared energy error versus computation time for integrations of the Sun and 8 planets of the Solar System. Each curve shows calculations using a particular integrator for various step sizes. $\sixa$ and $\sixc$ are different 6th-order algorithms given by Eqns.~\ref{eq_s6a} and \ref{eq_s6c} respectively.}
\end{figure}

Figure~4 shows the performance of this algorithm compared to $S_{6A}$ for the same problem shown in Figures~2 and 3. In this case, the efficiency of the new 6th-order algorithm $\sixc$ is only a modest improvement over that of $\sixa$, presumably because applying the analytic corrector is still somewhat expensive. The algorithm does appear to behave as a 6th-order integrator for large step sizes. However, the error quickly saturates to a constant level at smaller step sizes due to the error incurred by the terms in $H_I$ and $H_S$, which are integrated with the same step size for all the cases in Figure~4. For this problem, at least, it appears that the method used by integrator $\sixb$ is more efficient than applying an analytic corrector as in $\sixc$.

%
%
\section{Round-Off Error}
For small step sizes, the main source of error in $N$-body integrations is ``round-off'' error caused by the limited precision of the computer rather than the accuracy of the integration algorithm. For T+V integrators, we can greatly reduce round-off error, at very little computational cost, using a simple procedure.

\begin{figure}
\includegraphics[width=\columnwidth]{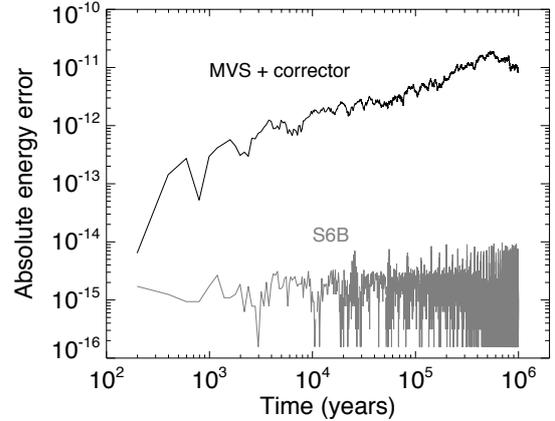}
\caption{Absolute energy error versus time for an integration of the Sun and 8 planets of the Solar System. The curves show integrations using the sixth-order T+V symplectic integrator $\sixb$ and a second-order MVS integrator with a symplectic corrector. The step size in each case is 0.23 days.}
\end{figure}

Consider a step in which the coordinates ${\bf X}$ are modified by an amount ${\bf dX}$. Typically, ${\bf dX}$ will be much smaller in magnitude than ${\bf X}$, especially for small step sizes. The information stored in the least significant digits of ${\bf dX}$ is lost when it is added to ${\bf X}$, and this error accumulates stochastically (or worse, depending on the computer) over the course of many steps. The same argument applies to changes in the momenta ${\bf P}$.

We can save much of this lost information by keeping track of the accumulated changes ${\bf dX}$ and ${\bf dP}$, updating these at each sub-step, and noting exactly how much of this information is transferred to ${\bf X}$ (or ${\bf P}$) when the quantities are updated. This is actually trivial to achieve in practice using the following procedure:
\begin{itemize}
\item At the start of an integration, set ${\bf dX}={\bf dP}=0$.
\item Every time a new ${\bf dX}$ or ${\bf dP}$ is calculated, add it to the existing value of ${\bf dX}$ or ${\bf dP}$.
\item When updating ${\bf X}$ (or ${\bf P}$), follow these steps:
\begin{itemize}
\item Store the coordinate values before the update ${\bf X}_0={\bf X}$.
\item Update the coordinates: ${\bf X}={\bf X}_0+{\bf dX}$.
\item Modify the changes: ${\bf dX}\rightarrow{\bf dX}+({\bf X}_0-{\bf X})$.
\end{itemize}
Note that the inclusion of the parentheses in the last step is essential for the procedure to work.
\end{itemize}

Following these steps often reduces round-off error by 2--3 decimal orders of magnitude, and requires minimal extra computational cost. Figure~5 shows an  example using a million-year integration of the Sun and the 8 planets of the solar system. The figure shows the absolute energy error versus time for the sixth-order T+V integrator $\sixb$, using the same step size for $I$ as the other parts of the Hamiltonian. The result is compared to MVS leapfrog with a symplectic corrector. The step size for both algorithms is 0.23 days.

The energy error increases over time for the MVS algorithm due to a combination of round-off error and errors incurred by the routine that advances the Kepler problem. After 1 million years, the error is roughly 1 part in $10^{11}$, and it is likely to increase further for longer integrations. By contrast, the energy error for the T+V algorithm remains very small, less than 1 part in $10^{14}$ throughout the integration. The error at the end of the simulation is only slightly larger than that after only 100 years. This suggests that T+V integrators may be preferred to MVS algorithms for long-term integrations for which a high degree of accuracy is required.

%
%
\section{Summary}
Symplectic integrators separate the Hamiltonian for an $N$-body system into two or more parts that can be solved easily in isolation. The evolution of the complete system is approximated by combining multiple sub-steps that alternately advance one of the sub-systems. For $N$-body systems with a dominant central body, a common strategy is to separate the problem into (i) Keplerian orbits about the central body, and (ii) interactions between the satellites \citep{wisdom:1991}. These algorithms are called mixed-variable symplectic (MVS) integrators. 

In this paper, we re-examine another class of symplectic integrators in which the Hamiltonian for the system is split into terms involving the kinetic energy $T$ and the potential energy $V$ respectively \citep{gladman:1991}. 

The main conclusions of this study are
\begin{enumerate}
\item Classical second and fourth-order T+V integrators require substantially more computer time than MVS algorithms for the same level of accuracy.
\item The speed of T+V integrators can be improved by using different step sizes for strong and weak forces associated with primary-satellite and satellite-satellite terms respectively.
\item More efficient fourth and sixth-order T+V algorithms can be developed using force gradients and symplectic correctors. We describe 3 new sixth-order algorithms that require either 2 or 3 force evaluations per step, plus force gradients and their derivatives.
\item The fourth and sixth-order integrators are often more efficient than a second-order MVS (leapfrog) integrator, and are competitive with a second-order MVS algorithm that includes a symplectic corrector.
\item High-order T+V algorithms like these may be especially favorable compared to second-order MVS for systems containing many planets (which reduces the fractional cost of the Keplerian motion and indirect terms).
\item Round-off errors for T+V integrators can be reduced greatly, at little extra computational cost, using a simple modification that keeps track of the accumulated changes in the coordinates and momenta.
\end{enumerate}

\section*{Acknowledgements}
I would like to thank an anonymous referee for helpful comments on this paper.


\bsp	
\label{lastpage}
\end{document}